%% file: main.tex
\documentclass[sigconf,natbib=true,anonymous=false]
{acmart}
\include{Preamble/packages}

\include{Preamble/copyright}
\include{Preamble/definitions}
\include{Preamble/authors}
\begin{document}

\title{Reconstructing Content with Collaborative Attention for Universal Multimodal Representation Learning}

\begin{abstract}
Multimodal embedding models, rooted in multimodal large language models (MLLMs), have yielded significant performance improvements across diverse tasks such as retrieval and classification. However, most existing approaches rely heavily on large-scale contrastive learning and offer limited exploration of how the architectural and training paradigms of MLLMs affect embedding quality.
While effective for generation, the causal attention and next-token prediction paradigm of MLLMs does not explicitly encourage the formation of globally compact representations, limiting their effectiveness as multimodal embedding backbones. To address this, we propose \textbf{\methodname}, a \textbf{Co}ntent reconstruction pre-training paradigm based on \textbf{Co}llaborative \textbf{A}ttention for universal multimodal representation learning. Specifically, we restructure the attention flow and introduce an EOS-based reconstruction task, encouraging the model to reconstruct input from the corresponding $\langle \mathrm{EOS} \rangle $ embeddings. This drives the multimodal model to compress the semantic information of the input into the $\langle \mathrm{EOS} \rangle $ token, laying the foundations for subsequent contrastive learning.
Extensive experiments on MMEB-V1 demonstrate that \methodname~built upon Qwen2-VL and Qwen2.5-VL significantly improves embedding quality. Results validate that content reconstruction serves as an effective strategy to maximize the value of existing data, enabling multimodal embedding models to generate compact and informative representations, raising their performance ceiling. Our project is available at \url {https://github.com/Trustworthy-Information-Access/CoCoA}.

\end{abstract}

%

\maketitle
\input{Sections/Intro}
\input{Sections/Related_work}
\input{Sections/Method}

\input{Sections/Experiments}

\input{Sections/Overall_performance}
\input{Sections/Analysis}

\input{Sections/Case_Study}
\input{Sections/Conclusion}
\section*{Acknowledgments}
This work was funded by New Generation Artificial Intelligence-National Science and Technology Major Project of No. 2025ZD0123301, the National Natural Science Foundation of China (NSFC) under Grants No. 62302486 and No. 62441229, the Innovation Project of ICT CAS under Grants No. E361140, and the Strategic Priority Research Program of the CAS under Grants No. XDB0680102.
\bibliographystyle{ACM-Reference-Format}
\balance
\bibliography{refs}

\end{document}

%% file: Preamble/packages.tex
\usepackage{multirow}
\usepackage{placeins}
\usepackage{graphicx}
\usepackage{subcaption}
\usepackage{float}  
\usepackage{enumitem}
\usepackage{pgfplots}
\usepackage{makecell}
\AtBeginDocument{%
  }

%% file: Preamble/copyright.tex
\setcopyright{acmlicensed}
\copyrightyear{2026}
\acmYear{2026}
\setcopyright{cc}
\setcctype{by}
\acmConference[SIGIR '26]{Proceedings of the 49th International ACM SIGIR Conference on Research and Development in Information Retrieval}{July 20--24, 2026}{Melbourne, VIC, Australia}
\acmBooktitle{Proceedings of the 49th International ACM SIGIR Conference on Research and Development in Information Retrieval (SIGIR '26), July 20--24, 2026, Melbourne, VIC, Australia}
\acmDOI{10.1145/3805712.3809695}
\acmISBN{979-8-4007-2599-9/2026/07}

\ccsdesc[500]{Information systems~Retrieval models and ranking}

%% file: Preamble/definitions.tex
\newcommand{\methodname}{CoCoA} 
\newcommand{\heading}[1]{\vspace*{0.5mm}\noindent\textbf{#1.}}

%% file: Preamble/authors.tex
\author{Jiahan Chen}
\affiliation{%
  \institution{State Key Laboratory of AI Safety
Institute of Computing Technology,
Chinese Academy of Sciences
University of Chinese Academy of
Sciences}
  \city{Beijing}
  \country{China}
}
\email{chenjiahan24s@ict.ac.cn}

\author{Da Li \\ Hengran Zhang}
\affiliation{%
  \institution{State Key Laboratory of AI Safety
Institute of Computing Technology,
Chinese Academy of Sciences
University of Chinese Academy of
Sciences}
  \city{Beijing}
  \country{China}
}
\email{{lida21s,zhanghengran22z}@ict.ac.cn}

\author{Yinqiong Cai \\ Lixin Su}
\affiliation{%
  \institution{Baidu Inc.}
  \city{Beijing}
  \country{China}
}
\email{{caiyinqiong,sulixin}@baidu.com}

\author{Jiafeng Guo}
\affiliation{%
  \institution{State Key Laboratory of AI Safety
Institute of Computing Technology,
Chinese Academy of Sciences
University of Chinese Academy of
Sciences}
  \city{Beijing}
  \country{China}
}
\email{guojiafeng@ict.ac.cn}

\author{Daiting Shi \\ Dawei Yin}
\affiliation{%
  \institution{Baidu Inc.}
  \city{Beijing}
  \country{China}
}
\email{{shidaiting01,yindawei02}@baidu.com}

\author{Keping Bi}
\authornote{Keping Bi is the corresponding author.}
\affiliation{%
  \institution{State Key Laboratory of AI Safety
Institute of Computing Technology,
Chinese Academy of Sciences
University of Chinese Academy of
Sciences}
  \city{Beijing}
  \country{China}
}
\email{bikeping@ict.ac.cn}
\renewcommand{\shortauthors}{Jiahan Chen et al.}

%% file: Sections/Intro.tex
\section{Introduction}
Multimodal embedding, a foundational paradigm in cross-modal learning, aims to project heterogeneous modality-specific inputs into a shared high-dimensional semantic space, where semantically equivalent content from different modalities is mapped to comparable representations~\citep{clip, mme5, gme, blip}. 
By bridging the inherent semantic gaps between heterogeneous data, this technique enables models to interpret and correlate information across modalities rather than in isolation.
As a cornerstone component, multimodal embedding underpins a spectrum of high-impact tasks, ranging from cross-modal retrieval (e.g., text-to-image and image-to-text retrieval for content discovery) to advanced vision-language understanding tasks such as visual question answering, visual commonsense reasoning, and multimodal sentiment analysis. Its practical value extends to real-world applications~\citep{lifeir, food101, fashioniq,copali}, including personalized content recommendation, intelligent retrieval and categorization of e-commerce product images, and natural language-enabled semantic query for medical image databases, cementing its role as a pivotal driver of progress in modern information systems.


Previous explorations, exemplified by methods such as CLIP~\citep{clip}, BLIP~\citep{blip}, ALBEF~\citep{albef}, and their variants~\citep{uniir,coca,fgclip,filip,flamingo}, constructed multimodal embeddings by first encoding heterogeneous modality inputs separately, then fusing features via cross-modal alignment. Recently, Multimodal Large Language Models (MLLMs), have made remarkable strides~\citep{llava,qwen25vl,minigpt}, demonstrating transformative progress in unifying multimodal perception, generation, and reasoning. The success of MLLMs has unlocked new directions for multimodal embedding models.
These MLLMs can be directly optimized as the backbone of multimodal embedding models. E5-V~\citep{e5-v} introduces a unimodal contrastive learning paradigm, training the language branch of MLLMs on sentence pairs to narrow cross-modal representation gaps. VLM2Vec~\citep{vlm2vec} presents the Massive Multimodal Embedding Benchmark (MMEB) comprising 36 datasets across 4 meta-tasks, and a contrastive training framework to transform state-of-the-art MLLMs into embedding models. Building on these foundations, LLaVE~\citep{llave} proposes a simple yet effective framework that dynamically enhances the embedding model’s representation of negative pairs based on their discriminative difficulty.

Under the influence of the scaling law~\citep{scaling_law}, existing work focuses on optimizing contrastive learning, concentrating on how to modify the training process to enable MLLM-based embedding models to learn effectively from massive data. However, several structural challenges remain insufficiently explored.  Firstly, \textbf{the causal attention mechanism in MLLMs is not appropriately designed for embedding-oriented objectives}. Mainstream multimodal embedding models~\citep{gme, vlm2vec} inherit causal attention from their MLLM backbones, where each token attends only to preceding tokens. While this unidirectional attention is essential for autoregressive generation, it does not explicitly encourage holistic semantic fusion across modalities, making it less suited for constructing compact and information-dense multimodal representations~\citep{llm2vec}.  Secondly, \textbf{there exists a fundamental mismatch between next-token prediction and contrastive learning}. Next token prediction prioritizes sequential coherence for generation, whereas contrastive learning aims to produce globally compact embeddings that capture the holistic semantic content of multimodal inputs. As a result, it is often necessary to employ alternative training objectives to bridge the gap between the two tasks. Several recent studies have proposed targeted strategies. 
UniME~\citep{unime} employs knowledge distillation to enable the text modules within MLLMs to learn from a competitive text embedding model. MoCa~\citep{moca} replaces the causal attention mechanism in MLLMs with a bidirectional attention mechanism. It further pre-trains the modified model on a massive corpus of 30B tokens, reorienting MLLMs from sequence generation toward embedding output, which imposes requirements on data scale.

In this work, we propose \textbf{\methodname}, an effective pre-training strategy that enables MLLMs to transition into competitive multimodal embedding models. 
We restructure the attention flow and introduce an EOS-based reconstruction task, which prompts the model to reconstruct the original input using the corresponding $\langle \mathrm{EOS} \rangle $ embeddings. This incentivizes the multimodal model to distill and compress the full semantic information of the input into the $\langle \mathrm{EOS} \rangle $ token, thereby laying a solid foundation for subsequent contrastive learning.
The end-to-end training process consists of three phases, which progressively transform causal attention-based MLLMs into effective multimodal embedding encoders. 
Firstly, we activate and warm up bidirectional attention through joint multimodal reconstruction, by performing masked next token prediction (MNTP) on text and masked autoencoding (MAE) on images, enabling unrestricted information flow across modalities and establishing a foundation for deep semantic fusion. 
Secondly, we introduce an EOS-Bridged compression mechanism via attention truncation. During encoding, inputs are structured into a compression-side and a reconstruction-side connected by a shared $\langle \mathrm{EOS} \rangle$ token. By aggressively masking the text-side tokens and optimizing MNTP on the text block conditioned on $\langle \mathrm{EOS} \rangle$, we force multimodal information from the visual block to be summarized and compressed into the $\langle \mathrm{EOS} \rangle$ token, yielding compact and information-dense embeddings. Finally, we perform contrastive learning based on the compressed embeddings, aligning multimodal representations in a unified semantic space. 

This staged design ensures that contrastive learning operates on compact and information-rich embeddings, rather than attempting to induce alignment directly from loosely organized representations. Experiments on MMEB-V1 demonstrate that \methodname~is both efficient and effective. In the comparison of multimodal embedding models with a small size ($\leq 3B$), it achieved state-of-the-art performance. In the comparison of $7B$ models, \methodname~achieved competitive results using a negligible amount of pre-training data compared to other methods, and still demonstrated an upward trend when additional data is introduced. We also conducted further ablation experiments to analyze how \methodname~works. Our results show that through content reconstruction, the potential of MLLMs in the field of multimodal embedding can be unlocked, thereby establishing a robust foundation for large-scale contrastive learning. In summary, our contributions are as follows:
\begin{itemize}[leftmargin=*, topsep=0pt, partopsep=0pt, parsep=0pt]
    \item We propose \methodname, an effective training paradigm to mitigate the mismatch between next token prediction and contrastive learning, by introducing the auxiliary task of content reconstruction to enable holistic multimodal fusion, compressing multimodal context into compact and information-rich embeddings.
     \item We compared \methodname with competitive multimodal embedding models. The results demonstrate that our approach achieves performance comparable to SOTA using significantly less data. This underscores the efficacy of \methodname. 
    \item We also conducted detailed experiments to analyse how \methodname works. The results demonstrate that it is the quality rather than the quantity of data that determines whether our reconstruction is effective and can aid in optimising contrastive learning.
\end{itemize}

%% file: Sections/Related_work.tex
\section{Related Work}
\subsection{Multimodal Embedding}
Multimodal embedding, as a core technology for cross-modal interaction, establishes a foundation for downstream tasks such as retrieval and classification by mapping information from different modalities, including text, images, and video, into a high-dimensional vector space, thereby enabling semantic matching~\citep{clip, blip, vista, vlm2vec, coma}. 

Early Vision-Language Models (VLMs) process information across different modalities through deep fusion, such as CLIP~\citep{clip} and BLIP~\citep{blip}, which achieve multimodal information interaction and deep fusion by independently encoding images and text and training under specific objectives. 
CLIP is pre-trained on a large corpus of weakly annotated image-text pairs, aligning visual and textual representations in a shared latent space through symmetric contrastive learning. BLIP introduces a multi-task pre-training framework based on CLIP, enhancing cross-modal representation performance in retrieval and classification tasks by unifying contrastive learning, image-text matching, and image captioning tasks. However, the deep fusion paradigm inherently suffers from insufficient cross-modal interaction, which limits its performance in fine-grained tasks. To mitigate this issue, VLMs have explored early fusion to strengthen cross-modal interactions between visual and textual modalities~\citep{vista,vlm2vec,gme}. For instance, VISTA~\citep{vista} improves the quality of multimodal embeddings by incorporating visual representations as input to text embedding models. 

Driven by the rapid adoption of multimodal large language models (MLLMs), multimodal embedding has undergone a pivotal paradigm shift. MLLMs can serve as powerful backbones for embedding models. There are several refined training strategies have been proposed to enhance the performance of MLLM-based embedding models. For example, VLM2Vec~\citep{vlm2vec} extends contrastive learning to a multi-task setting to improve robustness across heterogeneous objectives. GME~\citep{gme} approaches the problem from a different perspective by synthesizing large-scale, high-quality fused-modal training data to enhance multimodal embedding learning. 
Contrastive learning is inherently coarse-grained, as it prioritizes instance-level similarity and provides limited supervision for fine-grained semantics matching. Moreover, this paradigm is highly sensitive to the quality and hardness of negative samples, where low-informative negatives can drastically hinder the discriminative power of learned representations. Consequently, recent work has focused on refining the contrastive learning paradigm. LLaVE~\citep{llave} further enhances alignment by introducing hardness-weighted contrastive learning together with cross-device negative sample gathering. UniME-V2~\citep{unime-v2} leverages MLLMs to mine hard negative samples, making embedding model training more efficient.


\subsection{Pretraining for Multimodal Embedding}
Contrastive learning, as a training paradigm for embedding optimization, has significantly advanced the matching capabilities of multimodal embeddings through global similarity constraints. However, it is hard to scale without labeled pair data~\citep{laion_400m, siglip_2}. Contrastive learning cannot leverage the vast amount of unpaired multimodal data available on the internet. Lack of diversity in training objectives leads to suboptimal cross-modal alignment~\citep{mme5}. Contrastive learning encourages alignment based primarily on surface-level similarity, rather than deeper semantic or contextual understanding. As a result, the learned embedding models often struggle to generalize to diverse multimodal scenarios. 
These limitations highlight the need for more expressive training objectives that explicitly facilitate semantic understanding over heterogeneous data distributions. To address this gap, early pioneering works, such as ALBEF~\citep{albef} and BLIP~\citep{blip}, integrate auxiliary supervision signals into multimodal pre-training pipelines. Specifically, they employ Image-Text Matching and Masked Token Prediction on text, which helps inject richer semantic supervision into multimodal representations. Recent work, UniME~\citep{unime}, proposes an additional pretraining stage to improve multimodal embedding quality by leveraging textual discriminative knowledge distillation, in which a competitive LLM-based embedding model serves as a teacher to enhance the language representations of the MLLMs. 
Beyond auxiliary objectives, some approaches further recognize the limitations of causal attention in VLMs for multimodal context fusion and
advocate enabling bidirectional attention during pretraining. MoCa~\citep{moca} introduces a modality-aware continual pretraining stage, converts the original causal VLM into a bidirectional multimodal encoder, and integrates Masked Language Modeling (MLM) and Masked Autoencoding (MAE), improving the model’s ability to capture bidirectional contextual dependencies and align multimodal semantics. 

%% file: Sections/Method.tex
\section{Preliminary}
\heading{Multimodal Embedding Models}
Multimodal embedding models encode images and text through two key steps: first, converting images and text into sequences of patches and tokens; then, using a unified backbone model to map these visual patches and text tokens into vectors within a shared latent space.
Application of diverse pooling strategies on yielded feature sequences results in a corresponding embedding for the input.
Recently, Multimodal Large Language Models (MLLMs), renowned for their robust cross-modal understanding abilities~\citep{llava, qwen25vl}, can serve as the backbone for multimodal embedding models. Most of them adopt causal attention, under which information from different modalities is aggregated unidirectionally, and embeddings are typically extracted from the last $\langle \mathrm{EOS} \rangle$ token~\citep{prompteol}.

\heading{Contrastive Learning}
Contrastive learning plays a central role in multimodal embedding optimization by enforcing semantic consistency across modalities. Its core objective is to pull semantically relevant query-target pairs closer in the unified embedding space while pushing apart irrelevant pairs. 
This training paradigm has been adopted in downstream tasks such as multi-modal retrieval, owing to its effectiveness in learning discriminative embeddings~\cite{gme,vlm2vec,vlm2vecv2}. Given an MLLM-based embedding model, we obtain the corresponding embedding by feeding the query and the target into the model. Specifically, we extract the embeddings $h_{q}$ and $h_{t^{+}}$ from the last layer hidden state of the $\langle \mathrm{EOS} \rangle$ token. To optimize the embedding space, we adopt the InfoNCE~\citep{infonce} objective with in-batch negative samples, formulated as:
\begin{equation*}
\mathcal{L}
= - \log
\frac{\exp(\mathrm{sim}(h_q, h_{t^{+}})/\tau)}
{\exp(\mathrm{sim}(h_q, h_{t^{+}})/\tau)
+ \sum_{t^{-} \in \mathcal{N}} \exp(\mathrm{sim}(h_q, h_{t^{-}})/\tau)},
\end{equation*}
where $\mathcal{N}$ denotes the set of all negative samples, $\mathrm{sim}(\cdot)$ denotes the similarity function and $\tau$ is a temperature hyperparameter.

\begin{figure*}[t]
    \centering
    \includegraphics[width=0.8\textwidth]{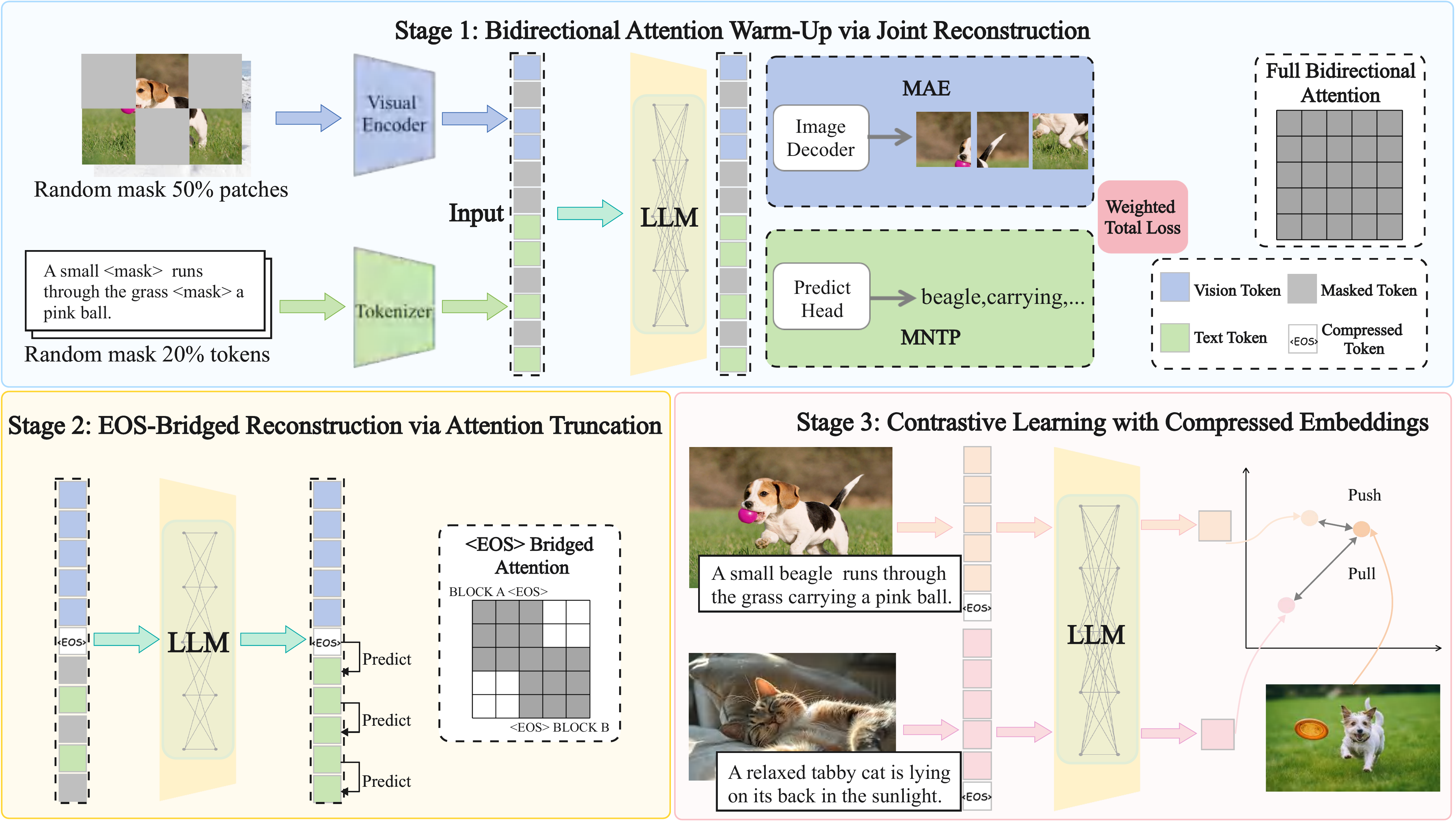}
    \caption{Overview of \methodname. Stage 1: Bidirectional attention warm-up using MAE and MNTP for different modalities; Stage 2: EOS-bridged caption or answer tokens reconstruction forces the multi-modalities' semantics to be compressed into a single token, i.e., $\langle \mathrm{EOS} \rangle$; Stage 3: Contrastive learning using $\langle \mathrm{EOS} \rangle$ .}
    \label{fig:placeholder}
\end{figure*}

\section{Method}

MLLMs can serve as the backbone for embedding models and have demonstrated remarkable performance in tasks such as retrieval and classification. However, the causal attention mechanism inherent to these models creates a bottleneck for cross-modal information fusion: it rigidly controls information access such that only tokens in subsequent positions can extract and utilize contextual information from tokens that appear earlier in the sequence. Moreover, without additional training, the representations obtained from MLLMs tend to be weakly organized, making direct contrastive learning insufficient for fully exploiting the training data to generate compact and semantically well-structured embeddings. To address the aforementioned bottleneck and empower embedding models to comprehensively capture and fuse cross-modal contextual information, we propose an end-to-end training process consisting of three stages. In the first stage, we employ bidirectional attention to replace the original causal attention in MLLMs, and we deploy joint sequence reconstruction tasks to guide the MLLMs' adaptation to bidirectional contextual dependency. And then we further exploit bidirectional attention to reconstruct the content for explicitly compressing the multimodal input, resulting in compact yet semantically rich high-quality embeddings. Finally, contrastive learning is performed using the compressed $\langle \mathrm{EOS} \rangle$ token.


\subsection{Bidirectional Attention Warm-Up via Joint Reconstruction}
Mainstream multimodal embedding models~\citep{vlm2vec, gme} inherit causal attention from their backbones. However, the restricted information flow may limit symmetric multimodal interaction and hinders the model’s ability to integrate visual and textual semantics fully. In large language models, prior work~\citep{llm2vec, llm_ql} have shown that modifying the attention mechanism can improve representation learning. However, directly replacing the causal attention in the backbone with bidirectional attention leads to suboptimal performance~\cite{llm2vec}. To adapt the model to bidirectional attention, LLM2Vec introduces a masked next token prediction (MNTP) task before contrastive learning. This pre-training step effectively initializes and activates the bidirectional attention mechanism, laying a foundation for subsequent contrastive learning. Motivated by this insight, we adopt a similar strategy in the first stage of our pretraining.

For the text modality, we adopt a masked next token prediction (MNTP) objective to warm up the bidirectional attention mechanism. Specifically, we randomly mask 20\% of the input text tokens and train the model to recover each masked token based on its surrounding context while preserving the autoregressive prediction paradigm. For a masked token at position $i$, the model is trained to predict it using the output at the position $i-1$. The MNTP loss is formulated as:
\begin{equation}
\mathcal{L}_{\text{MNTP}}
= - \sum_{i \in \mathcal{T}_{\text{mask}}}
\log p_{\theta}\!\left(
x_i \mid \mathbf{h}_{i-1}
\right),
\end{equation}
where $\mathcal{T}_{\text{mask}}$ denotes the set of masked text positions, 
$x_i$ is the ground-truth token at position $i$, and $\mathbf{h}_{i-1}$ is the hidden representation at the previous position.

For the image modality, we draw inspiration from the MAE~\citep{mae}. Given an input sequence of image patches, we additionally sample a subset of image patches and replace them with Gaussian noise. 
Follow MAE, a shallow Transformer is employed to reconstruct pixel values within the masked image patch from its output yielded by MLLMs. The MAE loss is defined as:
\begin{equation}
\mathcal{L}_{\text{MAE}}
= \frac{1}{|\mathcal{P}_{\text{mask}}|}
\sum_{i \in \mathcal{P}_{\text{mask}}}
\mathrm{MSE}\!\left(
\hat{\mathbf{v}}_i,\ \mathbf{v}_i
\right),
\end{equation}
where $\mathcal{P}_{\text{mask}}$ denotes the set of masked image patches,
$\mathbf{v}_i$ and $\hat{\mathbf{v}}_i$ represent the ground-truth and reconstructed visual features of the $i$-th patch, respectively, and $\mathrm{MSE}(\cdot)$ denotes the mean squared error.
During the warm-up stage, the joint reconstruction objective is formulated as a weighted sum of the MNTP loss on the text modality and the MAE loss on the image modality, which is close to MoCa~\citep{moca}:
\begin{equation}
\mathcal{L}_{\text{warm-up}} = \mathcal{L}_{\text{MNTP}} + w 
\mathcal{L}_{\text{MAE}}. 
\end{equation}

\subsection{EOS-Bridged Reconstruction via Attention Truncation}
Benefiting from the bidirectional attention warm-up in the prior stage, which lifts the constraint imposed by causal attention, MLLMs can effectively integrate information from multiple modalities and capture cross-modal semantic relationships, enabling more efficient semantic extraction. However, relying solely on such global semantic extraction capabilities still fails to meet the core requirement of precise cross-modal semantic alignment in multimodal tasks. To bridge this gap, we introduce a second-stage pre-training specifically tailored to enhance the cross-modal alignment of MLLMs under the bidirectional attention mechanism. 
We divide a whole input sequence into two blocks and propose attention truncation to regulate cross-block information flow explicitly. 
The pre-training data in this stage follows a unified structural format: ``$\mathit{Block\ A} \, \langle \mathrm{EOS} \rangle \, \mathit{Block\ B}$'', where $\mathit{Block\ A}$ contains visual or multimodal inputs and $\mathit{Block\ B}$ includes only textual information. These two blocks are designed for different objectives. $\mathit{Block\ A}$ aims to extract and compress the corresponding content of its input. $\mathit{Block\ B}$ is used to reconstruct the original information using the embedding obtained from $\mathit{Block\ A}$. 
To explicitly control cross-block information flow, we introduce a special $\langle \mathrm{EOS} \rangle$ token that serves as the sole communication bridge between two blocks. Bidirectional attention is permitted within each block. It is also allowed between $\mathit{Block\ A}$ and $\langle \mathrm{EOS} \rangle$, as well as between $\langle \mathrm{EOS} \rangle$ and $\mathit{Block\ B}$. However, direct attention is strictly prohibited between $\mathit{Block\ A}$ and $\mathit{Block\ B}$, preventing the two blocks from accessing each other's representations. As a result, any cross-block information must be transmitted exclusively through the $\langle \mathrm{EOS} \rangle$, which is thereby forced to aggregate and compress multimodal contextual information.

Under the attention truncation constraint, the model is optimized using a masked next token prediction (MNTP) objective on $\mathit{Block\ B}$, where predictions are conditioned solely on the $\langle \mathrm{EOS} \rangle$ token and the remaining unmasked local context. This design intentionally limits the available textual context, thereby forcing the visual information from $\mathit{Block\ A}$ to be compactly encoded into the $\langle \mathrm{EOS} \rangle$ token. We apply an aggressive token masking strategy to $\mathit{Block\ B}$. Specifically, when the length of $\mathit{Block\ B}$ is fewer than four tokens, the entire sequence is masked; otherwise, \textbf{70}\% of its tokens are randomly masked. 
As a result, the model can successfully reconstruct the masked tokens in $\mathit{Block\ B}$ only if $\langle \mathrm{EOS} \rangle$ aggregates and summarizes the relevant multimodal content from $\mathit{Block\ A}$. Through this aggressive masking strategy, accurate token reconstruction becomes possible only when the $\langle \mathrm{EOS} \rangle$ representation captures high-level semantic information that bridges the visual and textual modalities. Consequently, the proposed attention truncation mechanism explicitly encourages the compression of multimodal information in $\mathit{Block\ A}$ into a compact and shared embedding, which is beneficial for subsequent contrastive learning.

\begin{table}[htbp!]
\caption{Construction of training instances in the form of
$\,\mathit{Block\ A}\, \langle \mathrm{EOS} \rangle \,\mathit{Block\ B}\,$ for different tasks during reconstruction.}
\centering
\resizebox{0.8\linewidth}{!}{
\begin{tabular}{lcc}
\toprule
\textbf{\textit{Meta-Task}} & \makecell[c]{\textbf{\textit{Block A}} \\ \textbf{\textit{(Compression)}}} & \makecell[c]{\textbf{\textit{Block B}} \\ \textbf{\textit{(Reconstruction)}}}  \\
\midrule
Classification & qry$_{img}$ & tgt$_{txt}$ \\
VQA & qry$_{img}$ + qry$_{txt}$ & tgt$_{txt}$ \\
Retrieval & qry$_{img}$ & tgt$_{txt}$ \\
\bottomrule
\end{tabular}}
\label{tab:task_format}
\end{table}

A core requirement of this pre-training paradigm is that content across $\mathit{Block\ A}$ and $\mathit{Block\ B}$ must be semantically related. 
We specifically adopt data instances from three meta-tasks in the train set of MMEB-V1: Classification, VQA, and Retrieval. 
In the training set of MMEB-V1, each instance is formalized as a query–target pair (abbreviated as qry–tgt). 
For different tasks, two blocks are constructed differently. For Classification and Retrieval, $\mathit{Block\ A}$ includes only the image, whereas $\mathit{Block\ B}$ is composed of the target caption or label. For VQA, $\mathit{Block\ A}$ consists of the visual input together with the query text, while $\mathit{Block\ B}$ contains the corresponding target answer. The input and output formats corresponding to different tasks are listed in Table~\ref{tab:task_format}.

\begin{figure}[htbp!]
    \centering
    \includegraphics[width=0.9\linewidth]{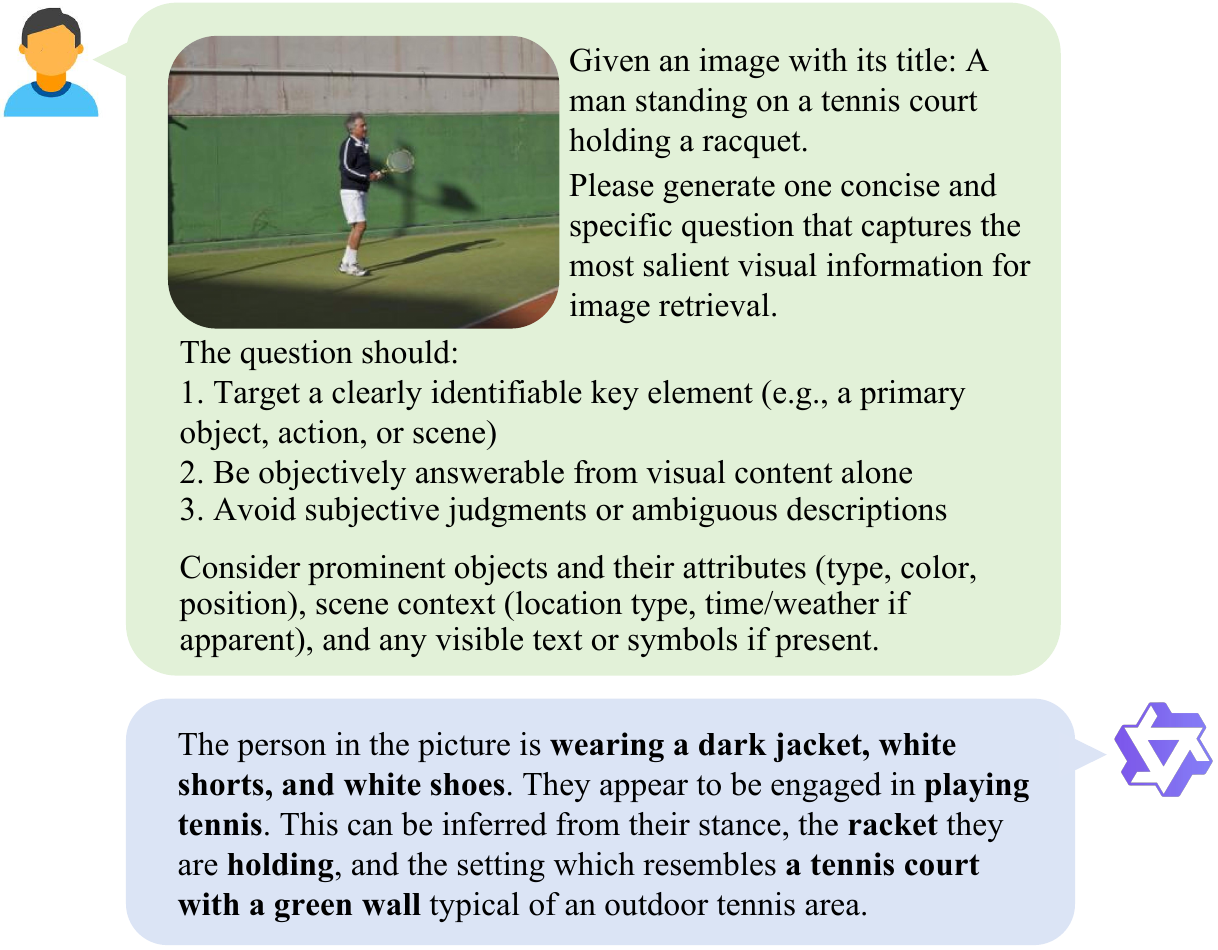}
    \caption{An example of synthetic data. The fine-grained details are highlighted in bold to emphasize that these elements are newly introduced by the synthetic data generation process and are not explicitly present in the original data.}
    \label{fig:data_aug}
\end{figure}

\heading{Synthetic Data Augmentation for Reconstruction Pretraining} 
Although the data sources in MMEB-V1 are diverse, they are collected from downstream tasks and lack semantic variation in their content. 
At the current stage, the semantic association and coverage across different modalities remain insufficient. 
This may reduce the difficulty of the reconstruction task, making it less effective in improving the quality of the embedding.
Therefore, we introduce synthetic data generated by MLLMs into our pretraining, leveraging their strong comprehension abilities to enrich the semantic diversity and quality of our data, like~\citep{qwen3vlemb}. We emphasize that the effectiveness of synthetic data in our method does not rely on model heterogeneity, but rather on its ability to integrate information across different modalities to generate richer and more diverse data, which is complementary to real downstream data.

As shown in Figure~\ref{fig:data_aug}, we generate two types of synthetic data from images in the MMEB-V1 trainset using the Qwen2.5-VL-7B model. The first type corresponds to the setting where the multimodal input contains only an image (i.e., $\mathit{Block\ A}$ consists solely of visual input). 
In this case, we prompt the model to produce a natural language description that captures the salient semantic content of the image, forming paired multimodal inputs for pretraining. 
The second type targets cross-modal scenarios where the multimodal input is composed of both an image and a question. We first prompt the model to generate questions that focus on specific attributes or semantic aspects of the image, and then require the model to produce corresponding answers conditioned on both the image and the generated question. This process results in multimodal samples that couple visual content with complementary linguistic supervision, enabling richer cross-modal semantic modeling. 

\begin{table*}[t]
\centering
\caption{Results on MMEB-V1. ``IND'' denotes in-distribution, and ``OOD'' refers to out-of-distribution. \textbf{Bold} and \underline{underline} indicate the optimal and suboptimal performance, respectively.}
\label{tab:mmeb}
\small
\renewcommand{\arraystretch}{0.9}
\resizebox{0.85\textwidth}{!}{
\begin{tabular}{l c c c c c c c c}
\toprule
\multirow{2}{*}{\textbf{Model}} & \multirow{2}{*}{\textbf{Size}} &
\multicolumn{4}{c}{\textbf{Per Meta-Task Score}} &
\multicolumn{3}{c}{\textbf{Average Score}} \\
\cmidrule(lr){3-6} \cmidrule(lr){7-9}
 &  & Classification & VQA & Retrieval & Grounding & IND & OOD & Overall \\
\midrule

\multicolumn{9}{l}{\textit{Only contrastive learning}} \\
CLIP (ViT-L)        & 428M & 55.2 & 19.7 & 53.2 & 62.2 & 47.6 & 42.8 & 45.4 \\
OpenCLIP (ViT-L) & 428M & 56.0 & 21.9 & 55.4 & 64.1 & 50.5 & 43.1 & 47.2 \\
GME (Qwen2-VL) & 2B & 56.9 & 41.2 & 67.8 & 53.4 & -- & -- & 55.8 \\
UNITE (Qwen2-VL) & 2B & 63.2 & 55.9 & 65.4 & 75.6 & 65.8 & 60.1 & 63.3 \\
VLM2Vec (Qwen2-VL) & 2B & 59.0 & 49.4 & 65.4 & 73.4 & 66.0 & 52.6 & 59.3 \\
VLM2Vec (Qwen2.5-VL)  & 3B   & 55.3 & 57.3 & 62.7 & 73.2 & -- & -- & 60.3 \\
E5-V (LLaVA-1.6)    & 7B & 39.7 &  10.8 & 39.4 & 60.2 &  34.2  &  33.9  & 33.9 \\
VLM2Vec (Qwen2-VL)  & 7B   & 62.6 & 57.8 & 69.9 & 81.7 & 72.2 & 57.8 & 65.8 \\
mmE5        & 11B  & \textbf{67.6} & 62.7 & 71.0 & \underline{89.7} & 72.4 & \underline{66.6} & 69.8 \\

\midrule
\multicolumn{9}{l}{\textit{Pretrain before contrastive learning}} \\
UniME (Phi3.5-V)        & 4.2B & 54.8 & 55.9 & 64.5 & 81.8 & 68.2 & 52.7 & 64.2 \\
UniME (LLaVA-1.6)        & 7B & 60.6 & 52.9 & 67.9 & 85.1 & 68.4 & 57.9 & 66.6 \\
MoCa (Qwen2.5-VL)        & 3B & 59.8 & 62.9 & 70.6 & 88.6 & 72.3 & 61.5 & 67.5 \\
MoCa (Qwen2.5-VL)        & 7B & \underline{65.8} & \underline{64.7} & \textbf{75.0} &92.4 & \underline{74.7} & \textbf{67.6} & \textbf{71.5} \\

\midrule
\multicolumn{9}{l}{\textit{Our method}} \\
CoCoA (Qwen2-VL) & 2B & 63.0 & 57.4 & 69.6 & 85.8 & 70.2 & 60.7 & 66.0 \\
CoCoA (Qwen2.5-VL) & 3B & 60.9 & 63.5 & 69.5 & 88.2 & 72.6 & 61.2 & 67.5 \\
CoCoA (Qwen2.5-VL) & 7B & 64.7 & \textbf{66.5} & \underline{72.5} & \underline{90.1} & \textbf{75.7} & 64.3 & \underline{70.6} \\

\bottomrule
\end{tabular}}
\end{table*}

\subsection{Contrastive Learning Based on Compressed Embeddings}
After the compression pretraining, we refine the multimodal representations through contrastive learning. For both the query and target inputs, we append the  $\langle \mathrm{EOS} \rangle$ at the end of the sequence and extract its last layer representation as the embedding used for contrastive learning. The $\langle \mathrm{EOS} \rangle$ token maintains methodological consistency across stages: having already compressed complex multimodal semantics into an information-dense representation through content reconstruction, it now provides the same high-level context for contrastive learning. Building upon this property, contrastive learning is applied directly to these compressed embeddings, encouraging semantically related query-target pairs to be pulled closer in the unified embedding space while pushing apart unrelated pairs. This design leads to more robust and discriminative multimodal embeddings, which are uniquely conducive for downstream tasks.

%% file: Sections/Experiments.tex
\section{Experimental Setup}
\subsection{Datasets}

\methodname~takes content reconstruction as its core design orientation. To precisely align with this core goal and avoid interference from irrelevant data, 
all data used for both pretraining and contrastive learning are drawn from the Massive Multimodal Embedding Benchmark (MMEB-V1), which comprises four meta-tasks: Classification, Visual Question Answering (VQA), Retrieval, and Visual Grounding, spanning a total of 36 datasets~\cite{A_OKVQA,Webqa,Objectnet,mscoco}. Among these, 20 datasets~\cite{docvqa,infographicvqa,n24news} are used for training. Across three training stages, we use the data sampled from these 20 datasets. 

\heading{Data for Pretraining} During the Bidirectional Attention Warm-Up via Joint Reconstruction Stage, we randomly sample 100K query-target pairs and train the MLLM for one epoch, which we find sufficient for stable convergence. In the following EOS-Bridged Compression via Attention Truncation Stage, we construct a more diverse pretraining corpus by combining both original and synthesized data. Specifically, we sample 300K instances from datasets belonging to three meta-tasks: Classification, VQA, and Retrieval. In addition, we further augment the pretraining data by generating approximately 200K high-quality multimodal instances using carefully designed prompts applied to images from the train set of MMEB-V1, thereby enriching the diversity and coverage of multimodal contexts. 

\heading{Data for Contrastive Learning} After completing the pretraining of reconstruction, we proceed to the contrastive learning stage, where only the original train set of MMEB-V1 is used, ensuring a fair evaluation setting without reliance on synthesized data.

\subsection{Implementation Details}
We adopt Qwen2-VL and Qwen2.5-VL as the backbone for \methodname. Our training pipeline consists of two major stages: pretraining for embedding compression and contrastive learning. The pretraining is further divided into bidirectional attention warm-up via joint reconstruction and EOS-Bridged compression via attention truncation. In the first pretraining stage, bidirectional attention is activated to warm up multimodal fusion. A masking ratio of 20\% is applied to text tokens for MNTP, while 50\% of the image patches are randomly masked for MAE. The loss weight for MAE is set to 0.5. During the second pretraining stage, bidirectional attention is retained while attention truncation is introduced to enforce visual-side compression. To this end, to preserve the integrity of visual information, no masking is applied to the multimodal input($\mathit{Block\ A}$) tokens in the stage. Instead, masking is performed exclusively on the reconstruction-side($\mathit{Block\ B}$): when fewer than four text tokens are present, all tokens are masked; otherwise, a random masking ratio of 70\% is applied. Under this asymmetric masking strategy, the model is trained with MNTP on the reconstruction-side, where the prediction of masked tokens is conditioned on the $\langle \mathrm{EOS} \rangle$ together with the limited remaining textual context. As a result, multimodal information from the multimodal input must be compactly encoded into the $\langle \mathrm{EOS} \rangle$, forcing the $\langle \mathrm{EOS} \rangle$ to act as a semantic bridge that aggregates and compresses visual context to support text reconstruction. Finally, in the third fine-tuning stage, the compressed multimodal embeddings are aligned through contrastive learning with a temperature parameter of 0.02, and the effective batch size is 1024. We adopt dynamic image resolution processing to handle multimodal inputs with varying image sizes. Across all training stages, we use LoRA (r=16), and all experiments are conducted on 8 A800 GPUs with a learning rate of 5e-5.

\subsection{Evaluation and Metrics}
We evaluate the performance of CoCoA on the eval set of MMEB-V1, a comprehensive benchmark that evaluates multimodal embeddings across four meta-tasks aligned with the training setting. It consists of 36 evaluation datasets, including 20 in-domain datasets and 16 out-of-domain datasets, enabling systematic assessment of both in-distribution and cross-domain generalization. We adopt Precision@1 as the evaluation metric.

%% file: Sections/Overall_performance.tex
\section{Experimental Results} 
We present the overall multimodal embedding performance on MMEB-V1 in Table~\ref{tab:mmeb}. 
For a rigorous comparative analysis, we select a diverse suite of state-of-the-art multimodal embedding models. Experimental results in Table~\ref{tab:mmeb} show that \methodname~achieves state-of-the-art performance in a comparison of methods on models of small scale ($\leq 3B$). 
In the comparison of $7B$ models, \methodname~achieved competitive results using a negligible amount of pre-training data compared to other methods. Results from \methodname~based on Qwen2-VL and Qwen2.5-VL demonstrate its generalizability.

These methods can be categorized into two distinct groups based on their training paradigms: those that exclusively leverage contrastive learning, and those that incorporate an additional pretraining phase before contrastive learning. Some findings can be drawn by comparing the performance of methods across different groups and by comparing the performance of methods within the same group: 
(1) Comparison of results between two categories shows that, under the same backbone, introducing a dedicated pretraining stage leads to notable improvements in embedding performance. 
(2) Comparisons among methods optimized solely through contrastive learning demonstrate the impact of the performance of MLLMs on multimodal embedding models.
(3) Compared with existing pretraining approaches, \methodname~achieves comparable results while using substantially less pretraining data and contrastive learning data. 
MoCa was pretrained with full parameters on a corpus containing 30B tokens, followed by contrastive learning on the MMEB-V1 training set and additional expanded data, totaling approximately 2 million query-target pairs. In contrast, \methodname~uses only 500K pretraining pairs, consists of 300K instances sampled from the train set of MMEB-V1 and an additional 200K high-quality synthetic pairs generated by Qwen2.5-VL-7B, and the whole train set of MMEB-V1 for contrastive learning. Our method achieves performance comparable to the best baseline, highlighting the effectiveness of explicit compression. 


%% file: Sections/Analysis.tex

\section{Further Analysis}
We conducted additional experiments to analyze how \methodname~works. To ensure fair and comparable analysis, all the experiments are conducted under a unified setting. 
We adopt Qwen2-VL-2B as the backbone and extract one million query-target pairs from MMEB-V1 for contrastive learning to enable efficiency in analysis.

\subsection{Ablation Study}
To demonstrate the value of each pretraining stage in \methodname, we compared performance changes after removing each stage individually from the existing end-to-end training. The same 100K pretraining samples are uniformly sampled from the train set of MMEB-V1 for this comparison. Results are shown in Table~\ref{tab:ablation}.
First, removing the EOS-Bridged Reconstruction via Attention Truncation leads to the most significant performance degradation. 
This observation highlights the central role of explicit content reconstruction: by reconstructing input content from the embedding of $\langle \mathrm{EOS} \rangle$, multimodal embedding models are compelled to compress input information into a compact embedding. It establishes a robust foundation for subsequent contrastive learning.
Second, eliminating the Bidirectional Attention Warm-Up stage also results in a clear performance drop. Without prior warm-up, the model struggles to exploit bidirectional dependencies, leading to suboptimal cross-modal fusion capabilities and poor compression performance, even when retaining the reconstruction task.


\begin{table}[htbp!]
\centering
\caption{Ablation study on different pretraining stages.}
\label{tab:ablation}
\resizebox{0.8\linewidth}{!}{
\begin{tabular}{lcr} \toprule
\multicolumn{2}{l}{\textbf{Model}}                                & \textbf{MMEB-V1}  \\ \midrule
\multicolumn{3}{l}{\textit{Our Method}}                                   \\
\addlinespace[0.2em]
\multicolumn{2}{l}{CoCoA (Qwen2-VL 2B)}                  & \textbf{62.9}  \\ \midrule
\multicolumn{3}{l}{\textit{Pretraining Stage Ablation}}             \\
\addlinespace[0.2em]
\multicolumn{2}{l}{\quad w/o. Bidirectional Attention Warm-Up} & 61.6      \\
\multicolumn{2}{l}{\quad w/o. EOS-Bridged Reconstruction}         & 60.7  \\ \midrule
\multicolumn{3}{l}{\textit{Baseline}}                                     \\
\addlinespace[0.2em]
\multicolumn{2}{l}{VLM2Vec (Qwen2-VL 2B)}                & 59.3  \\ 
\bottomrule
\end{tabular}}
\end{table}

We further examine the design of the warm-up stage. We compared the performance of models trained using different loss functions after subsequent training. The results are presented in Table~\ref{tab:loss_ablation}. 

\begin{table}[htbp!]
\centering
\caption{Impact of different loss functions in Bidirectional Attention Warm-Up.}
\label{tab:loss_ablation}
\resizebox{0.6\linewidth}{!}{
\begin{tabular}{ccc} \toprule
\textbf{Image Loss} & \textbf{Text Loss} & \textbf{MMEB-V1}     \\ \midrule
MAE        &MNTP       & 62.9 \\ 
MAE        &--         & 61.8 \\
--         & MNTP      & 62.3 \\
--         & MLM       & 57.9  \\ \bottomrule
\end{tabular}}
\end{table}

We can observe that removing either the loss function corresponding to the image or to the text alone will impair performance. This confirms that reconstruction objectives on both the textual and visual modalities are essential for establishing robust bidirectional fusion. Text-side MNTP encourages semantic dependency modeling across modalities, while image-side MAE promotes holistic visual understanding. Together, they enable the model to effectively leverage bidirectional attention for multimodal representation learning. 
The task of MLM requires that mask tokens be predicted through their contextual environment, which typically involves bidirectional information access. This task format differs from next token prediction and cannot leverage the causal flow of information within MLLMs. MNTP addresses these problems by combining next token prediction with masked language modelling~\citep{llm2vec}. Notably, in decoder-only MLLMs, text tokens are only exposed to unidirectional information flow during training, whereas visual tokens are inherently modeled with bidirectional attention; therefore, introducing bidirectional modeling on the text side leads to larger relative gains. In addition, many downstream tasks rely more heavily on textual semantics, further amplifying this effect.
\subsection{Influence of Mask Ratios}
\begin{figure}[htbp!]
    \centering
    \includegraphics[width=0.9\linewidth]{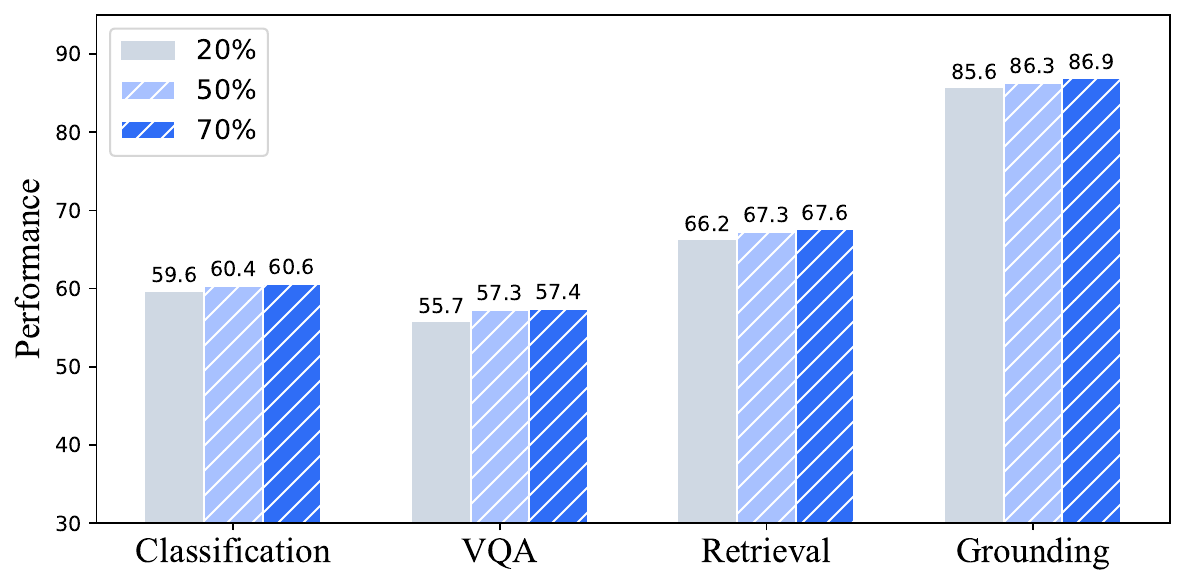}
    \caption{Performance under different mask ratios. We adopt three different mask ratios: 20\%, 50\%, and 70\% from left to right.}
    \label{fig:mask_rate}
\end{figure}

During the EOS-Bridged Reconstruction via Attention Truncation stage, we employ a masking strategy on the $\mathit{Block\ B}$ to facilitate semantic compression. The tokens in the $\mathit{Block\ B}$ are masked with a probability of 70\%. Specifically, when the $\mathit{Block\ B}$ contains fewer than four tokens, all of them are masked. In this part, we analyzed the impact of mask probability on performance. We conduct experiments under three masking ratios of 20\%, 50\%, or 70\%. And these experiments are all conducted on 300K samples sampled from the train set of MMEB-V1. We show the performance variation under different masking ratios in Figure~\ref{fig:mask_rate}.

We observe that as the mask ratio increases, the performance of the multimodal embedding models gradually improves. This trend clearly demonstrates the effectiveness of the reconstruction task: by reconstructing the content from the corresponding embeddings of $\langle \mathrm{EOS} \rangle$, the model is forced to aggregate and encode multimodal information into the $\langle \mathrm{EOS} \rangle$ token, resulting in a compact and information-rich embedding. However, there are subtle differences in the gains across different tasks. The performance gains observed for tasks of retrieval and VQA exceed those seen in classification. 
We attribute this difference to the varying lengths of text sequences across different tasks. We analyzed the distribution of text token lengths across different tasks. The corresponding statistical results are shown in Table~\ref{tab:token_stats}. 

\begin{table}[htbp!]
\centering
\caption{Token length statistics of reconstructed text across different tasks.}
\label{tab:token_stats}
\begin{tabular}{lcrrr}
\toprule
\textbf{Task} & \textbf{Avg. Tokens} & \multicolumn{1}{c}{\textbf{<4}} & \multicolumn{1}{c}{\textbf{4--10}} & \multicolumn{1}{c}{\textbf{>10}} \\
\midrule
Classification  & 2.41 & 89.3\% & 9.1\% & 1.6\% \\
VQA  & 3.12 & 67.8\% & 28.4\% & 3.8\% \\
Retrieval & 18.09 & 0.3\% & 20.1\% & 79.6\% \\
\bottomrule
\end{tabular}
\end{table}
For classification, the reconstruction content is typically short, and thus nearly all text tokens are masked regardless of the nominal masking ratio. For retrieval and VQA, where inputs requiring reconstruction are generally longer, increasing the masking ratio yields sustained performance gains. When handling long text, high masking ratios suppress reliance on residual textual context, compelling the model to reconstruct masked tokens based on the information compressed in the $\langle \mathrm{EOS} \rangle$ token. As a result, the embedding of $\langle \mathrm{EOS} \rangle$ becomes increasingly information-dense, establishing a solid foundation for contrastive learning. This indicates that content reconstruction requires the coordination of masking strategies to improve the quality of embeddings.


\subsection{Scaling of Pretraining Data}
\begin{figure}[htbp!]
    \centering
    \includegraphics[width=\linewidth]{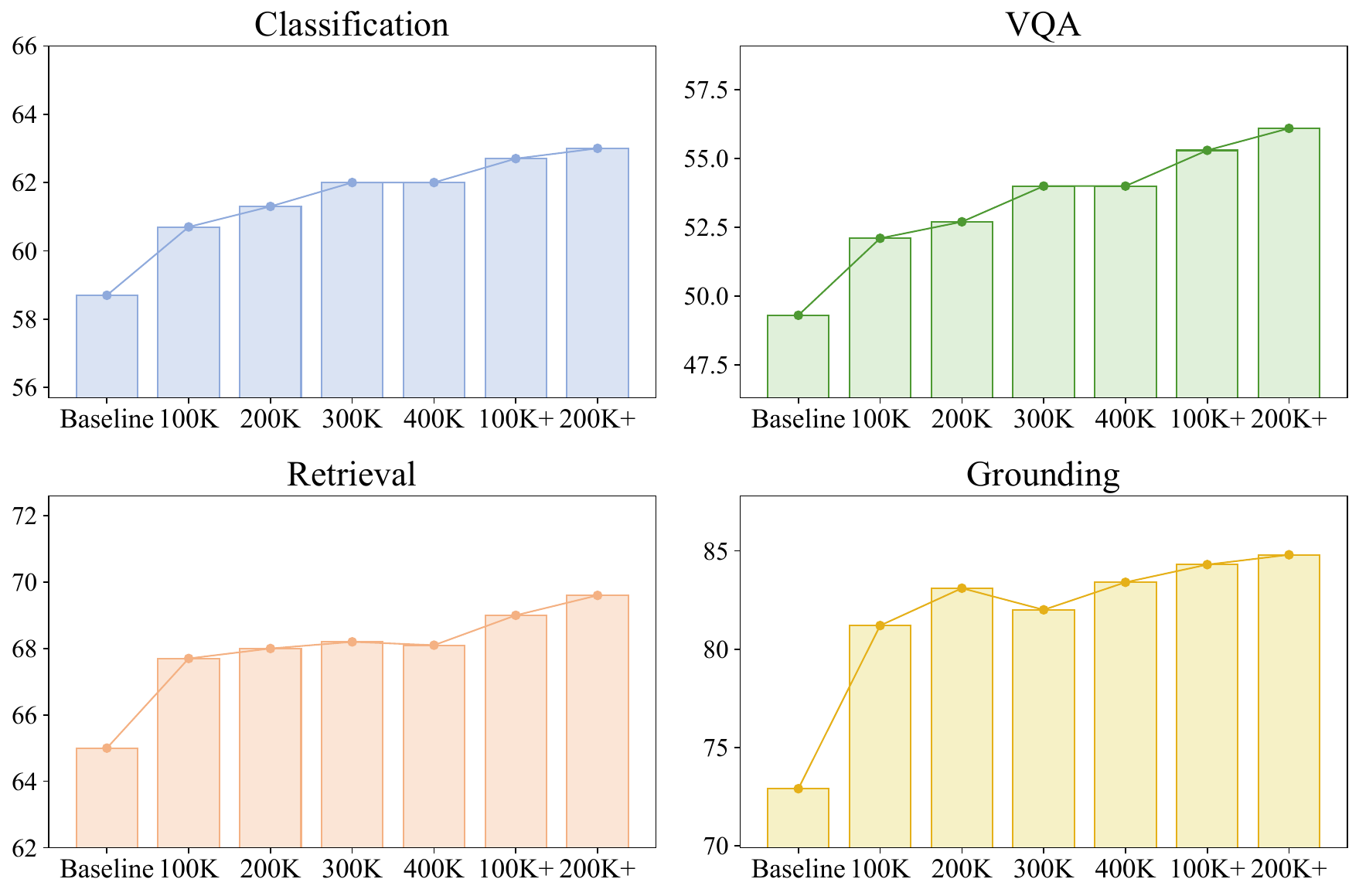}
    \caption{Performance under different pre-training data volumes. Baseline is VLM2Vec based on Qwen2-VL-2B. 100K-400K denote the amounts of MMEB-V1 original data, and ``+'' indicates additional synthetic data.}
    \label{fig:data_scale}
\end{figure}

\methodname~proposes an efficient pretraining strategy for multimodal embedding models, which relies on pretraining data that is not only inexpensive to construct but also widely available in abundant quantities. To analyze the impact of pretraining dataset scale on \methodname's performance, we vary the number of pretraining samples uniformly sampled from the train set of MMEB-V1, and then examine the effect of incorporating additional synthesized data. Specifically, we pretrain the model using 100K, 200K, and 300K samples drawn from MMEB-V1. We demonstrate the performance variations across different data scales in Figure~\ref{fig:data_scale}.

As shown in the figure, increasing the amount of pretraining data from 100K to 300K consistently improves performance across all tasks. This trend indicates that larger pretraining corpora provide richer multimodal information, enabling the model to better organize and compress multimodal information during the pretraining stage. However, when further increasing the pretraining data scale from 300K to 400K samples (still drawn from the train set of MMEB-V1), we observe no significant performance gains, with most meta-task scores remaining unchanged. We attribute this saturation effect to the limited diversity of MMEB-V1 itself: this portion of data was collected from specific tasks, and they tend to be semantically redundant.

To increase data diversity, we extend the train subset of MMEB-V1 by incorporating additional high-quality multimodal data synthesized by Qwen2.5-VL-7B. 
We observed that performance has been further improved across multiple tasks. This suggests that data diversity and semantic novelty, rather than sheer data volume alone, play a crucial role in pretraining. Synthetic data introduces richer semantic information, making reconstruction tasks more challenging. Consequently, even when the overall dataset is already substantial, it can still yield performance gains.

\begin{figure}[htbp!]
    \centering
    \includegraphics[width=0.9\linewidth]{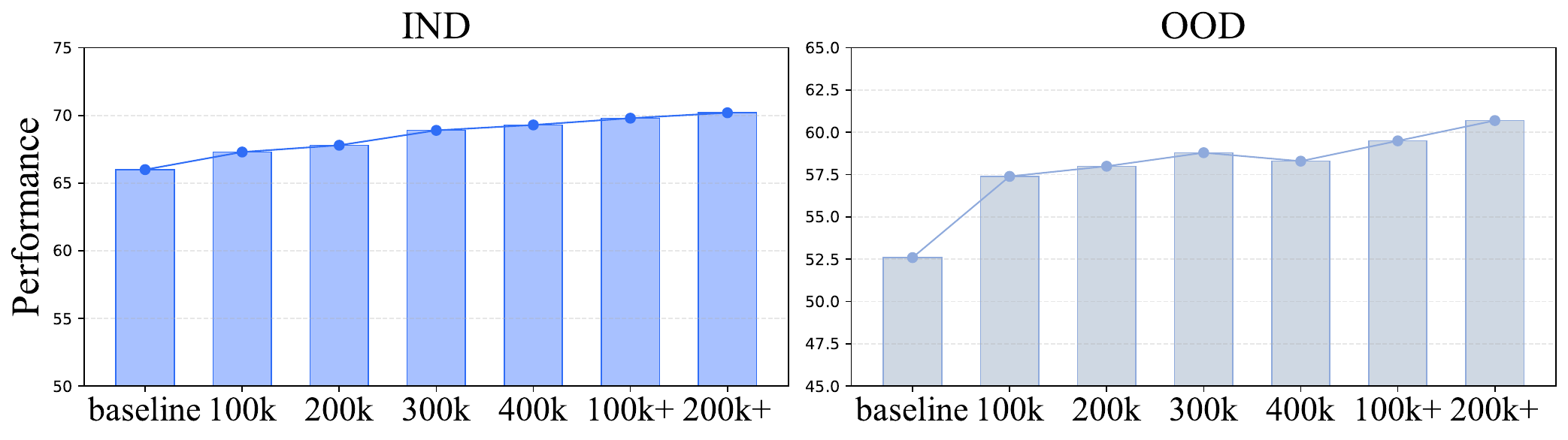}
    \caption{The effect of pretraining data scale on In-Domain (IND) and Out-of-Domain (OOD) performance.}
    \label{fig:ind_ood}
\end{figure}

In addition, we analyze the impact of pretraining data scale in both in-domain (IND) and out-of-domain (OOD) contexts. As the amount of original data from the MMEB-V1 training set increases, we observe consistent performance improvements on both IND and OOD performance up to 300K samples, indicating that moderate scaling of real data effectively enhances representation learning and generalization. 
However, when the data scale is further increased from 300K to 400K, the performance gain on IND begins to slow down, while OOD performance exhibits a noticeable decline. This suggests that simply increasing the amount of data within the same distribution may gradually introduce redundancy and bias the learned embeddings toward dominant in-domain patterns, thereby weakening generalization to unseen domains. After introducing synthetic data on top of the 300K original samples, both IND and OOD performance continue to improve. The additional synthetic data provides more diverse semantic coverage beyond the original data, which helps enhance robustness and generalization. Nevertheless, the performance gains gradually plateau as more synthetic data is added, indicating diminishing returns from data scaling once sufficient semantic diversity has been achieved.


%% file: Sections/Case_Study.tex
\section{Case Study}
To gain an intuitive understanding of how the compression pretraining stage organizes and structures multimodal embeddings, we conduct a qualitative study on representative examples. Specifically, we investigate whether the semantic information compressed into the $\langle \mathrm{EOS} \rangle$ token during pretraining is sufficient to support text reconstruction. We compare models with and without compression pretraining and using the $\langle \mathrm{EOS} \rangle$ representations to generate text tokens under the same decoding constraints.

\begin{figure}[htbp!]
    \centering
    \includegraphics[width=0.95\linewidth]{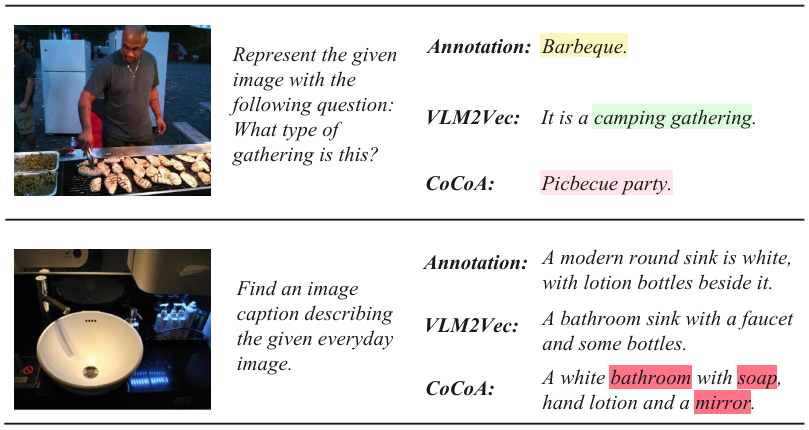}
    \caption{Case study of visual information reconstruction from $\langle \mathrm{EOS} \rangle$ in different multimodal embedding models.}
    \label{fig:case_study}
\end{figure}

As illustrated in Figure~\ref{fig:case_study}, \methodname~can preserve and recover fine-grained visual semantics more accurately. In the first example, our method effectively compresses and reconstructs high-level scene semantics from the visual input. Specifically, the model correctly captures the barbecue-related context in the image and recovers the semantic concept of a ``picbecue party''. By comparison, VLM2Vec produces a more generic and less precise description, ``It is a camping gathering.'', which overlooks the key visual information related to the barbecue scene.

The second instance illustrates an example where \methodname~produces an incorrect reconstruction content. From the reconstructed content, it can be seen that the retrieved semantics emphasize visual aspects that differ from the ground truth focus. In such cases, a single caption is insufficient to fully capture the rich content of the image. Although concepts such as ``soap'', ``bathroom'', and ``mirror'' are correctly inferred from the image, they introduce semantic distractions during retrieval, as they correspond to secondary visual elements rather than the primary retrieval intent. This suggests that when images contain multiple objects and diverse information, compressing the visual content from a single aspect may bias the representation toward certain aspects while suppressing others. These observations indicate that incorporating more comprehensive modeling could be critical for handling images with complex and multi-aspect visual information.


%% file: Sections/Conclusion.tex
\section{Conclusion and Future Work}
In this work, we present~\methodname, a pretraining task of reconstructing content via collaborative attention for the optimization of multimodal embedding. By employing a bidirectional attention mechanism and performing EOS-Bridged on the embedding obtained under truncated attention before contrastive learning, \methodname~compresses multimodal inputs into a compact and information-rich embedding. 
This design substantially improves the efficiency with which training data is utilized, allowing MLLM-based embedding models to achieve strong performance with significantly less pretraining and contrastive learning data. Extensive experiments and analysis demonstrate that \methodname~can unlock the potential of MLLMs as embedding models. Our case studies reveal that compressing visual information from a single embedding may be insufficient for complex images that contain multiple objects, attributes, or viewpoints. This observation suggests one promising direction is to explore a compression and reconstruction mechanism based on multiple embeddings, where different tokens capture complementary aspects of the multimodal content. 

 